\documentclass[twocolumn]{aastex631}
\usepackage[utf8]{inputenc}
\usepackage{CJK}
\usepackage{acronym}
\usepackage{xcolor}
\usepackage{ulem}
\usepackage{mathtools}
\usepackage{amsmath}
\usepackage{amssymb}
\usepackage{appendix}
\usepackage{float}
\usepackage{hyperref}
\usepackage{array} 
\usepackage{booktabs} 
\usepackage{tabularx}
\usepackage{romannum}
\usepackage{soul}

\newcommand{\msun}{\mathrm{M_\odot}}
\newcommand{\rsun}{\mathrm{R_\odot}}

\newcommand{\zsun}{\mathrm{Z_\odot}}

\newcommand{\ace}{$\alpha_\mathrm{CE}$}

\acrodef{ce}[CE]{common-envelope}

\begin{document}
\pagenumbering{arabic}
\begin{CJK*}{UTF8}{ipxm}
\title{Fits for the convective envelope mass in massive stars }

\author[0000-0002-8748-5236]{Lewis Picker}
\affiliation{School of Physics and Astronomy, Monash University, VIC 3800, Australia}

\author[0000-0002-8032-8174]{Ryosuke Hirai (平井遼介)}
\affiliation{School of Physics and Astronomy, Monash University, VIC 3800, Australia}
\affiliation{OzGrav: The Australian Research Council Centre of Excellence for Gravitational Wave Discovery, Clayton, VIC 3800, Australia}

\author[0000-0002-6134-8946]{Ilya Mandel}
\affiliation{School of Physics and Astronomy, Monash University, VIC 3800, Australia}
\affiliation{OzGrav: The Australian Research Council Centre of Excellence for Gravitational Wave Discovery, Clayton, VIC 3800, Australia}

\begin{abstract}
    We explore the evolution of massive stars ($>8\msun$) with 1-D models and present analytical fits to the masses and binding energies of the convective portions of their envelopes.  These fits are given as functions of total mass, metallicity, and surface temperature (used as a proxy for evolutionary phase).  They enable the application of the two-stage common envelope formalism \citep{ryo22} in rapid binary population synthesis frameworks.   We estimate that the degree of orbital hardening following common-envelope ejection spans 6 orders of magnitude and is a very strong function of the accretor mass, and, to a lesser extent, donor evolutionary phase. 
\end{abstract}

\section{Introduction}\label{sec:introduction}
\Ac{ce} evolution is a crucial, yet poorly understood phase in binary evolution \citep[see][for a detailed overview of \acp{ce}]{Iva13,10.1088/2514-3433/abb6f0}. It begins when one of the stars in the binary expands to overfill its Roche lobe, initiating dynamically unstable mass transfer onto its companion. The donated non-corotating material engulfs the binary and forms a \ac{ce}.  The companion experiences frictional forces within the \ac{ce} that cause it to spiral in towards the donor's core. Significant amounts of orbital energy are released and partly deposited into the expanding \ac{ce}, which may lead to its ejection. Once the \ac{ce} is ejected, the spiral-in will stop, leaving the core of the donor and the companion in a significantly tightened orbit.

\citet{pac76} originally introduced this idea as a potential formation pathway to justify the observed cataclysmic variables. Currently, the \ac{ce} scenario is applied to both high- and low-mass binaries as a potential pathway to form a wide range of interacting binaries such as type Ia supernova progenitors, X-ray binaries, and gravitational-wave sources. Studies involving \ac{ce} evolution share the primary goal of predicting the post-\ac{ce} state of the binary given the known properties of the pre-\ac{ce} system.

The complexity of the \ac{ce} phase and the insufficient understanding of the physics underlying the process lead to the adoption of an energy-based parameterisation of the \ac{ce} phase for binary population synthesis studies.  This approach, commonly referred to as the $\alpha$-formalism, relies on accounting for the energy budget, and thus implicitly treats the \ac{ce} event on a dynamical timescale, where radiative losses can be ignored.  This prescription allows users to predict the final separation of the post-\ac{ce} binary from known pre-\ac{ce} binary properties, given a suitable choice for the efficiency parameter ($\alpha_{\mathrm{CE}}$). This simple model is quick and ideal for use in population synthesis studies; however, there is growing evidence that $\alpha_{\mathrm{CE}}$ varies across systems, which casts doubt on the utility of the formalism \citep{Politano2004,Iaconi2019}. For example, \citet{wil22} report a qualitative difference between high-mass and low-mass donors and attribute it to the role of convection in the envelope, which calls for the use of different $\alpha_{\mathrm{CE}}$ values for different mass regimes.

Alternate formalisms for treating the \ac{ce} phase based on angular momentum loss \citep{gamma1,gamma2,2023ApJ...944...87D} have also been suggested. However, since the orbital separation of a binary can be reduced by orders of magnitude after a \ac{ce} phase and hence the remaining angular momentum is only a small fraction of the initial reservoir, the final separation becomes extremely sensitive to the model parameters of the formalisms. Therefore, any small variation in these parameters can lead to drastically different conclusions, severely limiting its predictive power. The $\alpha$-formalism is less sensitive to this issue by construction; on the other hand, it is very sensitive to the choice of location for the inner boundary of the envelope, which significantly impacts the envelope binding energy and hence the final separation \citep[e.g.,][]{Ivanova2011,Kruckow:2016,10.1093/mnras/stac237}. 

\citet{ryo22} recently proposed a two-stage formalism as a more physically motivated framework for treating \ac{ce} evolution in massive star binaries. They separate the whole \ac{ce} phase into two stages: adiabatic and non-adiabatic. Initially the mass transfer is unstable and hence a rapid inspiral ensues on a dynamical timescale, which can be treated with the $\alpha$-formalism. As the companion spirals deep into the star, the inspiral should slow down and transition to a slower more stable mass transfer on the thermal timescale, where energy generation and losses can no longer be ignored; instead, this stage can be treated through angular momentum balance. \citet{ryo22} suggest that this transition occurs roughly at the boundary of the convective and radiative portions of the donor envelope, inspired by findings from detailed 1D stellar modelling \citep{Fragos+2019,Marchant+2021,10.1093/mnras/stac237}. Although this is a relatively simple and well-defined boundary in 1D stellar models, this information is not readily available in rapid binary population synthesis codes where the stellar evolution is prescription-based. This brings forth the need to model the extra quantities required for this formalism: the mass and binding energy of the convective envelope.

In this paper, we present fitting formulae to predict the quantities required for the two-stage \ac{ce} formalism. In Section \ref{sec:method}, we describe in more detail the two-stage formalism and the construction of stellar models using the publicly available stellar evolution code MESA \citep{Paxton2019}. The results of our stellar models and fits are explained in Section \ref{sec:results}. We discuss the application and significance of the two-stage formalism in Section \ref{sec:discussion} and conclude this study in Section \ref{sec:conclusion}.

\section{Method}\label{sec:method}

\subsection{The Two-Stage CE formalism}\label{sec:2stage}
As the name suggests, the two-stage formalism of \citet{ryo22} separates a \ac{ce} event into two stages.  It segregates the donor star into three sections, namely the star's core, the radiative intershell, and the outer convective envelope. As such, the star's total mass can be summed up as

\begin{equation}
\label{eqn:star_mass}
 M_\mathrm{d} = M_\mathrm{core} + M_\mathrm{inter} + M_\mathrm{conv}, 
\end{equation}
where $M_\mathrm{d}$, $M_\mathrm{core}$, $M_\mathrm{inter}$, $M_\mathrm{conv}$ are the donor star's total, core, radiative intershell and convective envelope masses, respectively.

The main idea behind this new formalism is that the convective and radiative regions of the envelope have different thermal structures and therefore should respond to mass loss differently. Convective regions generally have flat or slightly negative entropy gradients, whereas radiative regions have positive entropy gradients. This difference is crucial as it relates to the dynamical response of the star's radius to mass loss \citep{soberman1997stability}. Massive red supergiants tend to have sizeable radiative intershell regions compared to lower-mass red giants where almost the entire envelope is convective. For this reason, \citet{ryo22} proposed that there should naturally be a qualitative difference between massive and low-mass star \acp{ce}.

Very roughly speaking, if mass were to be stripped from the convective region, it would expose a layer with higher entropy which would expand out from its original location. Mass transfer in this case is dynamically unstable and will likely result in a \ac{ce} event. On the other hand, on mass loss from a radiative region, the stellar radius would instead contract due to mass loss, preventing dynamically unstable mass transfer. However, the star would still be out of thermal equilibrium and would begin to expand on a thermal timescale instead \citep{10.1093/mnras/stac237}. In this case, mass transfer can be stable and can occur on a thermal timescale. In general, the stability of mass transfer is not as clear-cut as explained here but depends on many other factors such as the binary mass ratio, core mass, mass transfer rate, angular momentum loss from the system, and the physics of superadiabatic layers \citep[e.g.][]{Woods-Ivanova2011,Passy+2012,Temmink2023,Willcox+2023}.

Some detailed 1D simulations suggest that the convective/radiative boundary is a good approximation for the dividing line between the adiabatic (dynamical timescale) and non-adiabatic (thermal timescale) stages of the \ac{ce} phase \citep{Marchant+2021}. These simulations do not include the effect of dynamical heating through shocks, etc., which in principle could affect the response of the donor star. However, 3D hydrodynamical simulations by our group show that there is very little heating in the layers interior to the current companion location during the rapid inspiral phase \citep[e.g.][]{Lau:2021,Lau_2022} and therefore such approximations appear justified.

The \ac{ce} phase may be preceded by an extended stable mass transfer phase. In that case, the envelope structure upon entering the dynamical inspiral phase may be significantly altered from the structure of the donor when it initially overflowed its Roche lobe. Such effects are not captured in the current form of the two-stage formalism.

The $\alpha$-formalism can be applied to modelling just the dynamically unstable \ac{ce} event by attributing the change in orbital energy to unbinding the convective envelope. It can be expressed as
\begin{eqnarray}
E_\mathrm{bind} &=& \alpha_\mathrm{ce}\Delta E_\mathrm{orb}, \label{eqn:alpha}\\
\Delta E_\mathrm{orb} &=& -\frac{GM_\mathrm{d}M_\mathrm{2}}{2a_\mathrm{i}} + \frac{GM_\mathrm{d,post}M_\mathrm{2}}{2a_\mathrm{f}}, \label{eqn:Eorb}
\end{eqnarray}
where the parameter $\alpha_{\mathrm{CE}}$ is introduced to account for additional energy sinks (e.g., kinetic energy of ejecta) and sources (e.g., recombination energy). In the absence of any extra energy sources, the efficiency should be $ \alpha_{\mathrm{CE}} \leq 1 $. Here, $E_{\mathrm{bind}}$ is the binding energy of the convective envelope, $G$ is the gravitational constant, $M_\mathrm{2}$ is the mass of the accretor, $M_\mathrm{d,post}=M_\mathrm{core}+M_\mathrm{inter}$ is the post-\ac{ce} mass of the donor, and $a_\mathrm{i}, a_\mathrm{f}$ are the initial and final separations of the binary respectively.
The convective envelope binding energy is calculated by integrating the specific energy over the convective envelope mass, and it is commonly parameterized as 
\begin{equation}
\label{eqn:lambda}
    E_\mathrm{bind} = \int_{M_\mathrm{d,post}}^{M_\mathrm{d}}\epsilon(m)dm = \frac{GM_\mathrm{d}M_\mathrm{conv}}{\lambda R_\mathrm{d}},
\end{equation}
where $\epsilon(m)$ is the specific energy of the envelope,  $R_\mathrm{d}$ is the stellar radius of the donor, and $\lambda$ is an order unity parameter. The envelope specific energy generally refers to the gravitational potential energy but can include any other sources of energy within the envelope, such as thermal energy and the recombination energy that is released when the ions in the envelope cool down and recombine with electrons. 
The release of recombination energy has been proposed as a useful agent to help unbind the convective envelope \citep{Ivanova+2015,Ivanova2018,Reichardt+2019,Lau:2021,Lau_2022}.

Once the outer convective envelope has been ejected, the radiative intershell is exposed.   \citet{ryo22} assume that stable mass transfer from this intershell occurs on a thermal timescale. In this case, the $\alpha$-formalism becomes inadequate, and an alternative formalism that considers angular momentum conservation is used in the second stage of the \ac{ce} event.  For simplicity, we assume here that the mass transfer of the radiative intershell is completely non-conservative, and that all matter escapes the system with the specific angular momentum of the accretor. The degree of binary hardening under these assumptions can be expressed as \citep{pos14}

\begin{equation}
\label{eqn:momentum}
\begin{split}
\frac{a_\mathrm{f}}{a_\mathrm{i}} &= \frac{M_\mathrm{core} + M_\mathrm{inter} + M_2}{M_\mathrm{core} + M_2} \\
&\quad\times \left( \frac{M_\mathrm{core}+M_\mathrm{inter}}{M_\mathrm{core}}\right)^2 \exp\left( -2\frac{M_\mathrm{inter}}{M_2}\right).
\end{split}
\end{equation}
In principle, some amount of accretion may occur during this phase, and other choices could be made regarding how much angular momentum is lost from the system. For example, a circumbinary disk could be formed after the dynamical inspiral phase, influencing the angular momentum loss of the binary through resonant interactions \citep{KashiSoker:2011, GagnierPejcha:2023, Wei+2023}.

\subsection{MESA Models}\label{sec:MESA}
Utilising the 1D stellar evolution code MESA \citep{Paxton2011}, we evolve a grid of stars with varying initial masses and metallicities. We simulate an initial mass range of $5~\msun$ to $30~\msun$ in $1~\msun$ steps, across four sets of initial metallicities: $Z=0.03$, $1\,\zsun(\equiv 0.0142)$, $0.1\,\zsun$, and $0.01\,\zsun$. The mass range is specifically chosen because massive stars are expected to have a radiative intershell, whereas the range of metallicities incorporates star formation at different epochs that are predominantly responsible for the systems we observe today. We keep all other parameters constant, namely a mixing length of $\alpha_\mathrm{mlt}=1.6$ for convection and a step overshooting parameter of $\alpha_\mathrm{ov}=0.2$. The initial mass fraction of helium is set to $Y=0.24 + 2Z$.  No wind mass loss was applied in these models, and the Ledoux criterion is used for determining convective stability. 

At every time step, we record the convective envelope mass and its associated  binding energies. The convective envelope is defined as the convective region adjacent to the stellar surface. This method simplifies the often complex structure of stars by segregating them into just three sections: the tightly bound core where the hydrogen abundance is $X\leq0.01$, the radiative intershell, and a convective envelope that extends to the surface.

The complete inlist and extra files used to run the MESA models can be found at \href{https://zenodo.org/record/8287959}{https://zenodo.org/record/8287959}.

\section{Results}\label{sec:results}
\subsection{Convective envelope mass}\label{sec:Mconv}
\begin{figure}
    \includegraphics[width = \linewidth]{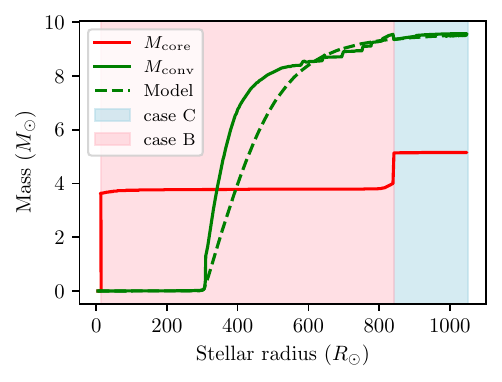}
    \caption{Masses of the helium core (red), the convective outer envelope (green), and the prediction of our model (dashed, green) are presented as a function of the stellar radius for a 15 $\msun$ star at solar metallicity. We show values for when the star reaches that radius for the first time, marking the moment when the donor may start transferring mass onto a companion. The rose-shaded region signifies case B mass transfer from a donor with an emerging convective envelope, while the blue region signifies a case C mass transfer. }
    \label{fig:15M_1Z_radius}
\end{figure}

Figure~\ref{fig:15M_1Z_radius} displays the mass evolution of the convective envelope and helium core as a 15$\msun$ star evolves from zero-age main sequence. We only show core and envelope masses at the time when the star first reaches a given radius, as this corresponds to the potential onset of a mass transfer event. During the main sequence phase, the stars in our mass range have convective cores and radiative envelopes with only a very thin convective zone close to the surface due to Fe and He ionization \citep[e.g.][]{Cantiello+2009}. Around $R_\mathrm{d}\approx 300~\rsun $, the surface layers become convectively unstable and the convective region deepens rapidly until the star reaches $R_\mathrm{d} \approx 500~\rsun$, and continues to grow slowly to its maximal value at the maximum radius ($R_\mathrm{d} \approx 1050~\rsun$). A non-negligible radiative intershell is present even at late stages in the star's evolution, as expected for massive stars. The jump in the core mass at $R \approx 20~\rsun$ in fact indicates the radius of the star upon core hydrogen exhaustion, which corresponds to when the core is born in our definition.  The rapid expansion of the star across the Hertzsprung gap leaves little time for the core mass to grow while the star expands up to $R_\mathrm{d}\approx 850~\rsun$. Beyond that radius there is another apparent steep jump in core mass, as the star only expands past that radius after helium is depleted in the core.  Importantly, the convective envelope mass connects smoothly across case B and case C mass transfer regimes, corresponding to donors before/during and after core helium burning, indicating that the stellar radius is closely correlated with the convective envelope mass, and the core mass plays only a minor role. This increase in the core mass comes at the expense of the radiative intershell.

\begin{figure}
    \centering
    \includegraphics[width = \linewidth]{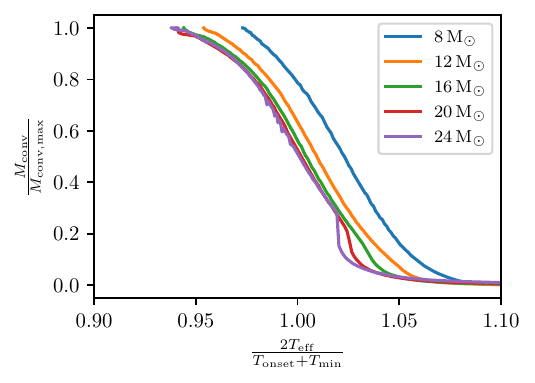}
    \caption{Normalised convective envelope mass vs.~normalised surface temperature for $0.1~\zsun$ stars, where $T_{\mathrm{onset}}=4400$~K.} 
    \label{fig:norm}
\end{figure}

In Figure~\ref{fig:norm}, we show the growth of the convective envelope against the effective temperature of the star for several masses. The convective envelope mass is normalised by the maximum value it reaches at the end of the evolution, while the temperature is normalised based on our model (see Section \ref{sec:modelling/Conv} for details). Compared to the sharp rise when plotted against the radius (Figure~\ref{fig:15M_1Z_radius}), the normalised convective envelope mass increases smoothly with the normalised surface temperature. Moreover, the shape of the curves appears to be self-similar over all masses, suggesting that normalised temperature may be a useful variable to fit against.

\subsection{Convective envelope binding energies}\label{sec:lambda}
\begin{figure}
    \centering
    \includegraphics[width = \linewidth]{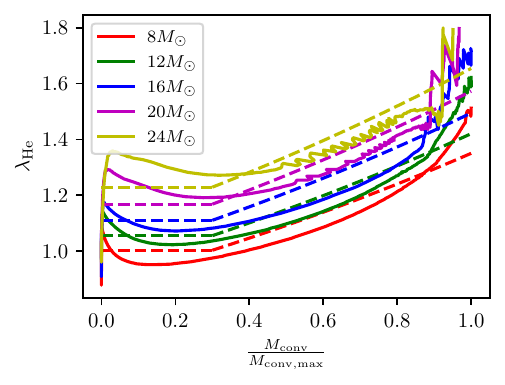}
    \caption{The effective $\lambda_\mathrm{He}$ values, including gravitational, thermal and helium recombination energy, are plotted on the ordinate, while the normalized convective envelope mass is plotted on the abscissa. Solid lines represent the calculated values of $\lambda_\mathrm{He}$, while dashed lines are the result of our fit. The different colours represent different masses within the $0.01~\zsun$ set.
}
    \label{fig:lambda_fit}
\end{figure}

Within our MESA simulations, we calculate the binding energies of the convective envelope while considering a number of energy components, specifically just the gravitational binding energy ($E_\mathrm{grav}$), gravitational + thermal ($E_\mathrm{bind}$), gravitational + thermal + He recombination ($E_\mathrm{He}$), and gravitational + thermal + total recombination ($E_\mathrm{total}$, which includes H and He recombination as well as molecular hydrogen formation). We put special emphasis on helium recombination energy even though it contains only $\sim$30 per cent of the total recombination energy, as its high ionisation temperatures makes it particularly significant for determining the post-inspiral separation \citep{Lau_2022}. For these four alternate binding energies, we calculate the equivalent value for $\lambda$ using Equation (\ref{eqn:lambda}). In Fig.~\ref{fig:lambda_fit}, we show the evolution of $\lambda_\mathrm{He}$ as a function of convective envelope mass. The behaviour of the curves is qualitatively similar over different masses. They start off with a value around $\sim$1 to 1.4, slightly declining up to $M_\mathrm{conv}/M_\mathrm{conv,max}\sim0.3$. Then they start to increase linearly with mass up to $M_\mathrm{conv}/M_\mathrm{conv,max}\sim0.9$ beyond which they oscillate. The later behaviour is related to thermal pulses observed in our models, where many of our assumptions break down (e.g., that of no mass loss). Similar overall behaviours are observed across different metallicities and the different definitions for binding energy.  For sets $Z = 0.03$ and $0.0142$, the $\lambda$ curves roughly overlap between $M_\mathrm{conv}/M_\mathrm{conv,max}\sim0.3$ and  $M_\mathrm{conv}/M_\mathrm{conv,max}\sim0.8$.

\subsection{Model fitting} \label{sec:modelling}
In this section we construct a set of fitting functions for the convective envelope mass and its associated binding energy values parameterised by $\lambda$. In doing so, we put special emphasis on choosing fitting parameters that are readily available in rapid population synthesis codes. 
\subsubsection{Mass fit \label{sec:modelling/Conv}}

\begin{figure}
    \centering
    \includegraphics[width = \linewidth]{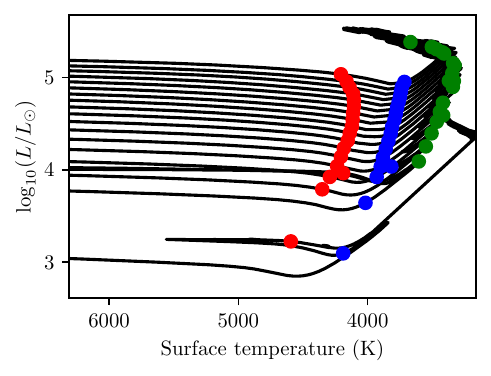}
    \caption{The evolutionary tracks of MESA stars in our $1\,\zsun$ dataset including $5,8-24\msun$, with markers that represent when the convective envelope mass reached 1\% (red), 50\% (blue), and 99\% (green) of its maximum.}
    \label{fig:conv_on_hr}
\end{figure}
\begin{figure}
    \centering
    \includegraphics[width = \linewidth]{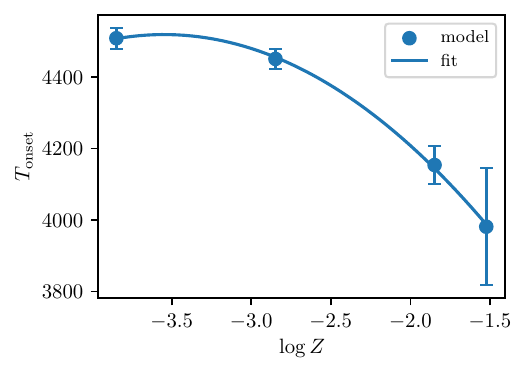}
    \caption{The median value of $T_\mathrm{onset}$ is plotted on the ordinate while logarithmic metallicity is plotted on the abscissa. The error bars represents the standard deviation between models with different stellar masses at the given metallicity.} 
    \label{fig:tonset_fit}
\end{figure}

Stars with the same metallicity begin to develop their convective envelope at roughly the same temperature, as seen in Fig.~\ref{fig:conv_on_hr}. Motivated by this behaviour, we introduce a parameter $T_{\mathrm{onset}}$, associated with the deepening of the convective envelope. $T_\mathrm{onset}$ is the surface temperature of the star when the convective envelope mass reaches 1\% of its maximum (red circles in Fig.~\ref{fig:conv_on_hr}). Although there is a slight mass dependence of $T_\mathrm{onset}$ in our models, we approximate it as a single value in our fits. We fit the metallicity dependence of $T_\mathrm{onset}$ with a quadratic fit in logarithmic metallicity: 
\begin{equation}
\label{eqn:Tonset}
T_\mathrm{onset} = -129.7\log(Z)^2 - 920.1\log(Z) + 2887.1,
\end{equation}
as shown in Fig.~\ref{fig:tonset_fit}. 
These and other fits against metallicity are not calibrated outside the range $Z\in[0.0001,0.03]$.
 Meanwhile, $T_\mathrm{min}$, the minimum surface temperature that stars reach just before carbon ignition, is a function of the stellar mass and metallicity that is already directly available in rapid stellar evolution models  \citep{hur00}.

Based on the normalised curves in Fig.~\ref{fig:norm}, we choose to fit the convective envelope mass as a function of surface temperature $T_\mathrm{eff}$:
\begin{equation}
\label{eqn:universal}
M_{\mathrm{conv}} = \frac{M_{\mathrm{conv,max}}}{1 + \exp\left(\frac{4.6(T_{\mathrm{min}} + T_{\mathrm{onset}} - 2T_{\mathrm{eff}})}{(T_{\mathrm{min}} - T_{\mathrm{onset}})}\right)}\,,
\end{equation}
where the constant $4.6$ empirically governs the slope of the convective envelope mass increase. Specifically,  it is determined by our definition for $T_\mathrm{onset}$, such that when $T_\mathrm{eff} = T_\mathrm{min}$ the mass of the convective envelope is 99\% of its maximum.  

\begin{figure}
    \centering
    \includegraphics[width = \linewidth]{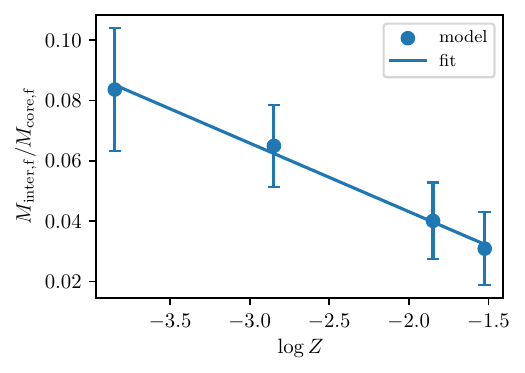}
    \caption{The median value of $M_\mathrm{inter,f}/M_\mathrm{core,f}$ is plotted on the ordinate, while the logarithmic metallicity is plotted on the abscissa.  The error bars represent the standard deviation between models with different stellar masses at the given metallicity.}
    \label{fig:gamma_fit}
\end{figure}

We create a fit for $M_\mathrm{conv,max}$ by noting that the ratio $M_\mathrm{inter,f}/M_\mathrm{core,f}$ of the final values of the intershell mass and the core mass at the onset of carbon burning roughly converges to a constant value within each metallicity data set. The exception to this are the $5\,\msun$ models, which do not have a permanent radiative layer. This ratio increases with decreasing metallicity, and thus we linearly model the median ratio with respect to logarithmic metallicity as 
\begin{equation}
\label{eqn:Minter}
M_\mathrm{inter,f}/M_\mathrm{core,f} = -0.023\log(Z) - 0.0023
\end{equation}
as shown in Fig.~\ref{fig:gamma_fit}.

Using this fit, we can fit for $M_\mathrm{conv,max}$ by re-writing Equation (\ref{eqn:star_mass}) as

\begin{equation}
    \label{eqn:gamma}
    M_\mathrm{{conv,max}} = \max\left(M_\mathrm{{d,f}} - M_\mathrm{{core,f}}\left(1 + \frac{M_\mathrm{{inter,f}}}{M_\mathrm{{core,f}}}\right), 0\right).
\end{equation}
Fig.~\ref{fig:15M_1Z_radius} displays the resulting fit along with the MESA model for a 15 $\msun$, 1~Z$_\odot$ star. 
    
\subsubsection{Binding energy fit}
The linear gradient (denoted $m_1$) of $\lambda$ with respect to the fractional convective envelope mass is relatively constant in the range between $0.3~M_\mathrm{{conv,max}}$ and $0.8~M_\mathrm{{conv,max}}$ (see Fig.~\ref{fig:lambda_fit}), so we use the average gradient of the $15\,\msun$ curve as a reference over metallicity. We use a constant approximation for $\lambda$ early in the post main sequence evolution, and avoid over-fitting the behaviour of $\lambda$ when MESA models become increasingly unstable beyond $\frac{M_\mathrm{conv}}{M_\mathrm{conv,max}} > 0.8$.  Thus, we fit $\ln{\lambda}$ via
\begin{equation}
\label{eqn:lnlambda}
\ln(\lambda) = 
  \begin{cases}
    m_1\frac{M_\mathrm{conv}}{M_\mathrm{conv,max}}+ b_1 & \mathrm{if}\ \frac{M_\mathrm{conv}}{M_\mathrm{conv,max}} > 0.3 \\
    0.3 m_1 + b_1 & \mathrm{otherwise}\ .
  \end{cases}
\end{equation}
In this fit, $b_1$ is itself a function of the initial stellar mass $M_\mathrm{i}$ given by 
\begin{equation}
\label{eqn:b1}
b_1 = m_2(M_\mathrm{i}/\msun) + b_2,
\end{equation}
where $m_2$ is given by 
\begin{equation}
\label{eqn:m2}
m_2 = a\log(Z)^2 + b\log(Z) + c. 
\end{equation}
The coefficients $a$, $b$, $c$, $m_1$ and $b_2$ are given in Table \ref{tab:lambda}.  Scipy's damped least squares algorithm (DLS) was used for all fits in this work \citep{2020SciPy-NMeth}.

\begin{table}[htbp]
    \caption{Coefficients for the $\lambda$ fits}
    \label{tab:lambda}
    \centering
    \begin{tabular}{@{}lccc@{}}
        \toprule
        Energy& [a,b,c]& $m_1$& $b_2$\\
        \midrule
        $E_\mathrm{grav}$ & [0.00048, -0.0017, -0.024] & -0.28 & -0.17 \\
        $E_\mathrm{bind}$ & [0.0018, 0.0061, 0.0067] & 0.35 & -0.22 \\
        $E_\mathrm{He}$ & [0.0023, 0.0088, 0.013] & 0.42 & -0.23 \\
        $E_\mathrm{tot}$ & [0.0032, 0.013, 0.026] & 0.65 & -0.27 \\
        \bottomrule
    \end{tabular}
\end{table}

\subsection{Summary of fit implementation}\label{sec:algorith_conv}
In summary, the following steps should be taken to implement our fit for the development of the convective envelope mass and associated values of $\lambda$:
\begin{enumerate}
    \item The parameters of our fits are available in machine-readable form \dataset[DOI: 10.5281/zenodo.8287959]{https://doi.org/10.5281/zenodo.8287959}.
    \item The core mass $M_\mathrm{{core,f}}$ and surface temperature $T_\mathrm{min}$ at the onset of carbon burning are already available in single stellar evolution fits \citep{hur00}, while $M_\mathrm{{d,f}}$ is set to the donor mass at the onset of the \Ac{ce} phase.
    \item Determine the ratio $M_\mathrm{inter,f}/M_\mathrm{core,f}$ using Equation~(\ref{eqn:Minter}).
    \item Determine $T_\mathrm{onset}$ using Equation~(\ref{eqn:Tonset}).
    \item Compute the maximum mass of the convective envelope $M_\mathrm{conv,max}$ using Equation~(\ref{eqn:gamma}).
    \item Compute the convective envelope mass $M_\mathrm{conv}$ as a function of surface temperature $T_\mathrm{eff}$ using Equation~(\ref{eqn:universal}).
    \item Calculate $m_2$ using Equation~(\ref{eqn:m2}) for the desired choice of binding energy (we use $\lambda_\mathrm{He}$ as a default), then calculate $b_1$ using Equation~(\ref{eqn:b1}).
    \item Calculate $\ln \lambda$ using Equation~(\ref{eqn:lnlambda}).
\end{enumerate}

Once the mass of the convective envelope and the associated $\lambda$ values are computed as above, the two-stage formalism can be used to predict the final separation after a \ac{ce} event as follows: 
\begin{enumerate}
    \item Calculate the radiative intershell mass $M_\mathrm{inter}$ using Equation~(\ref{eqn:star_mass}).
    \item Calculate the separation after stage 1 of the \ac{ce} event by computing the convective envelope binding energy $E_\mathrm{bind}$ using Equation~(\ref{eqn:lambda}), substituting the result into Equation~(\ref{eqn:alpha}) to compute the orbital energy $\Delta E_\mathrm{orb}$ using $\alpha_\mathrm{ce}=1$ as a default and inverting this to obtain the separation using Equation~(\ref{eqn:Eorb}).
    \item If the convective envelope is 'ejected', stage 2 of the \ac{ce} event begins. Use the final separation calculated at the end of stage 1 as the initial separation within Equation~(\ref{eqn:momentum}) to calculate the final separation between the donor's core and its companion.
\end{enumerate}

\section{Discussion}\label{sec:discussion}

\subsection{Fit limitations}
\begin{figure}
    \centering
    \includegraphics[width = \linewidth]{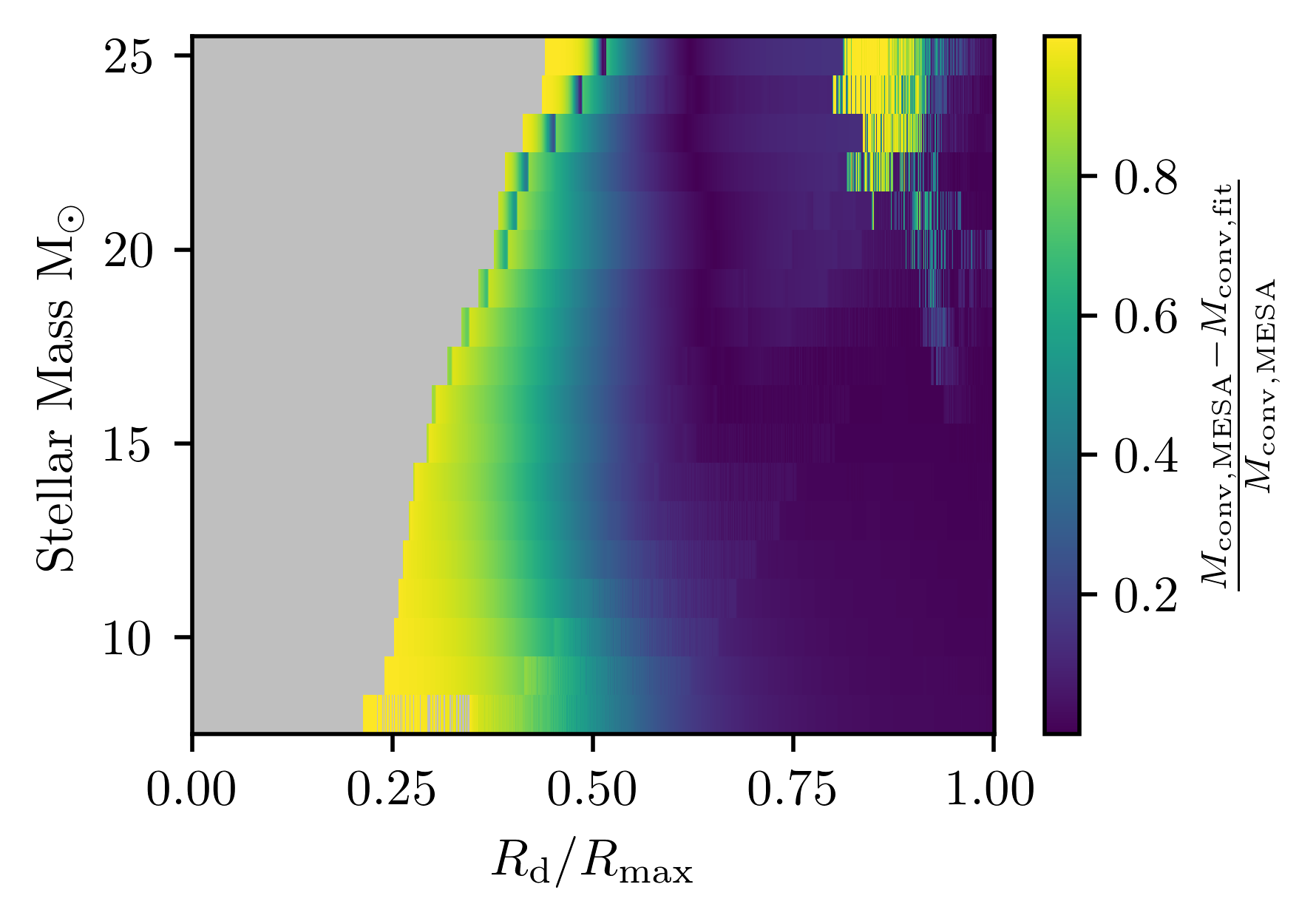}
    \caption{A 2D color map for $Z=\zsun$ displaying the fractional difference in the convective envelope mass between our fits and the MESA models with the stellar mass on the ordinate and the normalised radius on the abscissa.  The region where $M_\mathrm{conv,MESA}$ $<0.1~M_\odot$ and a significant convective envelope has yet to develop is shown in grey.  The artefact in the top right is a result of thermal pulsations.}
    \label{fig:fractional_error}
\end{figure}

Our fits made several simplifying assumptions that reduce their accuracy. For example, we approximate $T_\mathrm{onset}$ and $M_\mathrm{inter,f}/M_\mathrm{core,f}$ as functions of metallicity only, independent of the stellar mass, while the choice of the sigmoid fit in Equation~(\ref{eqn:universal}) imposes symmetry around its midpoint, where the gradient is highest. The accuracy of the resulting fits is shown in Figure~\ref{fig:fractional_error}, which displays the mismatch between the fits and the MESA models. The fits are generally accurate to $\sim 10\%$ in most of the region of interest, except where a convective envelope is only starting to develop (and hence the absolute accuracy of the fits is acceptable despite large fractional differences) or in the region where MESA models experience stellar pulsations.  As indicated in Figure \ref{fig:fractional_error}, our fits were predominantly calibrated to stellar models in the mass range 8--24 M$_\odot$, so applying these fits outside that range constitutes extrapolation with concomitant risks.

The MESA models themselves have systematic uncertainties; in particular, the uncertainty in the choice of mixing parameters. Stellar models that use higher mixing lengths will have more efficient convection and as a result expand less, leading to larger values of $T_\mathrm{min}$.
The choice of mixing parameters influences the existence of blue loops. In our models, we observe blue loops in the lower-mass, lower-metallicity models. Our assumption that the convective mass is strongly correlated with radius is slightly misguided during a blue loop phase. We observe that the convective mass decreases slightly ($\sim10~$per cent) after the star shrinks and re-expands back to its original size. However, the binding energy of the envelope exhibits a smooth transition back to its pre-loop values.

In these simulations, we explored the convective envelope mass and binding energy of evolved stars that did not lose mass.  In nature, stars in binaries will lose mass through winds and may lose or gain mass through stable mass transfer prior to the onset of a \ac{ce} phase. Or conversely, the donor star could have experienced mass gain in the past, which is known to influence the envelope binding energy \citep{Renzo+2023}. We find that the minimal mass of the radiative intershell is relatively insensitive to mass transfer history by sampling a few choices of mass loss rates and times, but further analysis is required to explore the mass loss/gain parameter space fully.

Despite all the above caveats, a point to note is that a high precision of the convective envelope mass and binding energy fits is not crucial in view of the overall accuracy of the two-stage formalism. As discussed in Section~\ref{sec:2stage}, the boundary between the adiabatic and non-adiabatic phases can depend on many other factors such as the stellar structure, details of inefficient convective energy transport, etc. There also could (should) be a marginally-adiabatic phase in between where the pure energy or angular momentum balance arguments do not hold. Therefore, despite the $\lesssim10$~per cent errors, the fitting functions we present in this paper provide satisfactory levels of accuracy in terms of the qualitative behaviour that the two-stage formalism aims to achieve.

\subsection{Effective \ace}\label{sec:effective}
\begin{figure}
    \centering
    \includegraphics[width = \linewidth]{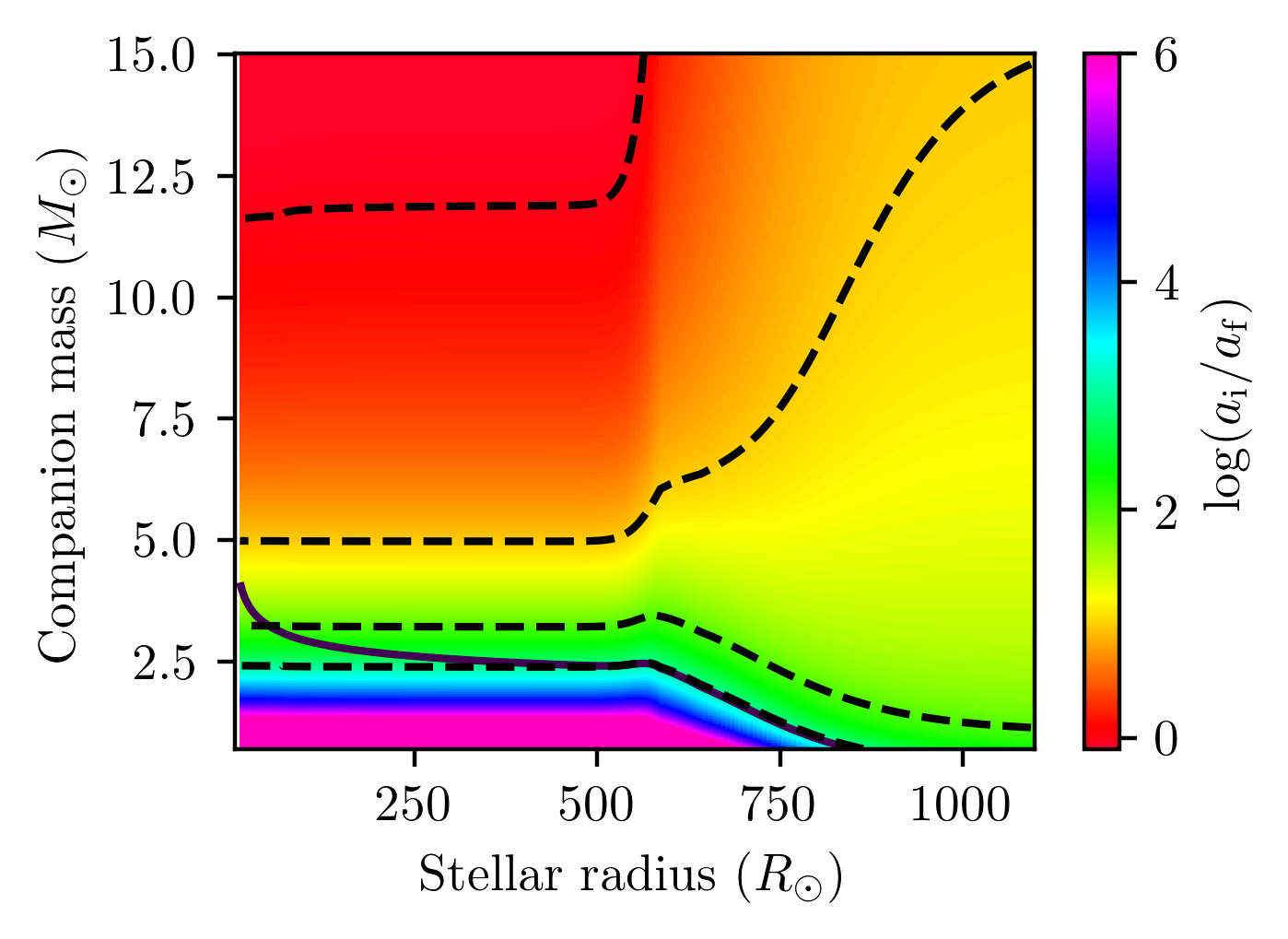}
    \caption{A 2D color map of the total orbital hardening as a function of donor radius at the onset of \ac{ce} and companion mass for a 15 $M_\odot$ solar metalicity donor. Dashed/dotted contours represents $\log(a_\mathrm{i}/a_\mathrm{f})=$ (0, 1, 2, 3) values from the top down. The solid curve represents a final separation of $a_f = 1\rsun$.}
    \label{fig:final_separation}
\end{figure}

To illustrate the consequences of the two-stage common envelope formalism we applied our fits to determine the final separation of binary systems experiencing \ac{ce} phases. We use our fits for $\lambda_\mathrm{He}$ and the stellar evolutionary tracks from COMPAS \citep{Riley_2022} based on \citet{hur00} stellar models. The initial separation $a_\mathrm{i}$ is that at which the donor of a given radius overfills its Roche lobe, whose size is calculated using the \citet{1983ApJ...268..368E} formula. We repeat this for a range of companion masses. Fig.~\ref{fig:final_separation} depicts a 2D map of the degree of binary hardening $\log(a_\mathrm{i}/a_\mathrm{f})$ for a $15~M_\odot$ solar metallicity donor star. Significant hardening only occurs for low-mass companions.  In this case, hardening is most extreme when the donor is early in its Hertzsprung gap evolution and the convective envelope mass is low, i.e., there is a lot of mass in the radiative intershell, leading the orbit to shrink by more than three orders of magnitude during the second, stable mass transfer stage.  On the other hand, if the companion's mass is close to that of the donor, the orbit will not shrink appreciably (in fact, can expand following mass ratio reversal) during the stage 2.  Consequently, massive companions can only experience moderate shrinking by about one order of magnitude during stage 1 of the \ac{ce} event when it happens late in the donor's evolution and a massive convective envelope has developed.

\begin{figure}
    \centering
    \includegraphics[width = \linewidth]{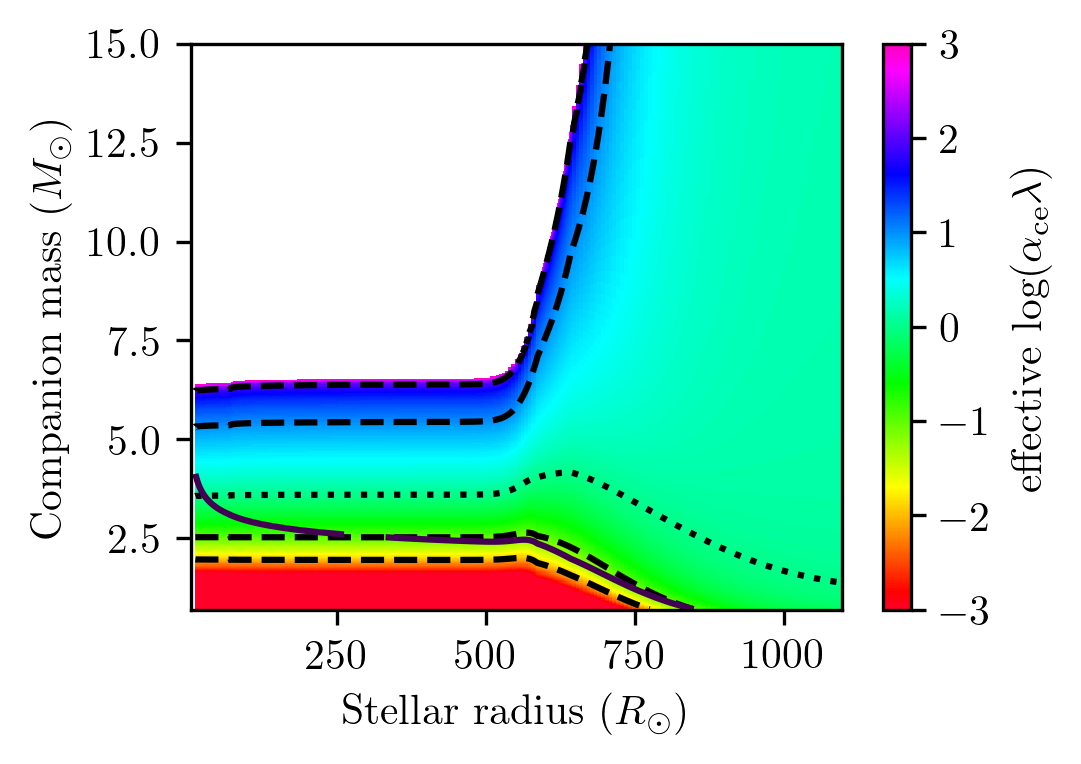}
    \caption{A 2D color map of the effective $\alpha_\mathrm{CE}\lambda$ parameter as a function of the donor radius and companion mass for a 15 $M_\odot$ solar metallicity donor. Dashed/dotted contours represent $\log(\alpha_{ce}\lambda)=$ (-2, -1, 0, 1, 2) values from top to bottom. The solid curve represents a final separation of $a_\mathrm{f} = 1\rsun$.}
    \label{fig:effective_alpha}
\end{figure}

Fig.~\ref{fig:effective_alpha} illustrates the $\alpha_\mathrm{ce}\lambda$ values that would be required in the traditional $\alpha$-formalism to reproduce the final separation obtained from the two-stage formalism. There is a very strong gradient in effective $\alpha_\mathrm{ce}\lambda$ values with companion mass, indicating that $\alpha_\mathrm{ce}$ should not be treated as a universal parameter in the traditional $\alpha$-formalism, if the two-stage prescription is accurate.  For moderately massive companions, $\alpha_\mathrm{ce}\lambda$ values are larger than unity (above the dotted contour), which implies that the two-stage formalism generally yields larger final separations than the traditional $\alpha$-formalism. The white region corresponds to the range of companion masses and donor radii for which $a_\mathrm{f}/a_\mathrm{i} > M_\mathrm{core}/M_\mathrm{d}$, making it impossible to replicate the final separation with the traditional $\alpha$-formalism. This occurs when the second, stable mass transfer stage is dominant due to the low convective envelope mass and fails to significantly shrink the binary due to the high companion mass. 
  
\subsection{Astrophysical consequences}\label{sec:Astrophysical consequences}
For massive companions, the two-stage formalism predicts wider post-\ac{ce} separations compared to the traditional $\alpha$-formalism, indicating that \ac{ce} evolution has a smaller role to play in the formation of binary black holes as sources of gravitational waves. However, less massive companions, such as neutron stars, can lead to binary hardening by several orders of magnitude during the \ac{ce} event, which can thus play a key role in the formation of tight double neutron stars, including those observed as radio pulsars in the Galaxy \citep[such as][]{1975ApJ...195L..51H} and as gravitational-wave events \citep[e.g.][]{GW170817}. The two-stage formalism may also help to explain how binaries with extreme mass ratios survive mass transfer, potentially addressing the puzzle of the formation of low-mass X-ray binaries \citep[e.g.][]{Podsiadlowski_2000, Podsiadlowski_2002}.

\section{Conclusion}\label{sec:conclusion}
We present fits for the convective envelope mass and binding energy of giant stars (with mass $>5 \msun$). The models demonstrate the presence of a significant radiative intershell layer throughout the star's post main-sequence evolution.  These fits make it possible to apply the two-stage \ac{ce} formalism of \citet{ryo22} in rapid binary population synthesis.
We leave detailed studies of the impact of the two-stage \ac{ce} formalism for future work, but show indicative results by combining our fits with COMPAS stellar tracks.  We find that the two-stage formalism corresponds to a very broad range of effective $\alpha_\mathrm{ce}\lambda$ values in the traditional $\alpha$-formalism, from \ace$\lambda\sim 10^{-3}$ to $10^3$. The stable mass transfer stage 2 has a dominant effect, with low-mass companions ($M_2 < M_\mathrm{core}$) leading to significant binary hardening. These results suggests that the findings of population synthesis studies that currently rely on a universal value of \ace\ may require significant revision.

\section*{Acknowledgments}
This work was performed on the OzSTAR national facility at Swinburne University of Technology. The OzSTAR programme receives funding in part from the Astronomy National Collaborative Research Infrastructure Strategy (NCRIS) allocation provided by the Australian Government. We thank Mike Lau for helpful comments on the manuscript.  RH \& IM acknowledge support from the Australian Research Council Centre of Excellence for Gravitational Wave Discovery (OzGrav), through project number CE230100016. IM is a recipient of the Australian Research Council Future Fellowship FT190100574.

\software{MESA \citep{Paxton2011,Paxton2013,Paxton2015,Paxton2018,Paxton2019},
          MESA SDK \citep{mesasdk},
          Matplotlib \citep{Hunter:2007},
          COMPAS \citep{COMPAS}
          }
\bibliography{ref}{}
\bibliographystyle{aasjournal}

\end{CJK*}
\end{document}